\documentclass[aps,prb,twocolumn,showpacs]{revtex4}
\usepackage{graphicx}
\usepackage{dcolumn}
\usepackage{bm}

\begin{document}

\title{Observation of vortex structure in MgB$_2$ single crystals by
Bitter decoration technique}

\author{L. Ya. Vinnikov}
\affiliation{Institute of Solid State Physics RAS, Chernogolovka,
Moscow region, 142432, Russia}
\author{J. Karpinski,  S. M. Kazakov and J. Jun}
\affiliation{ETH - Z\"{u}rich (H\"{o}nggerberg), HPF F-22, CH -
8093 Z\"{u}rich, Switzerland}
\author{J. Anderegg, S. L. Bud'ko, P. C. Canfield}
\affiliation{Ames Laboratory and Department of Physics and
Astronomy, Iowa State University, Ames, IA 50011, USA}

\date{\today}

\begin{abstract}
We report the observation of superconducting vortices in pure and
lightly Al doped MgB$_2$ single crystals. Low field experiments
allow for the estimation of the London penetration depth, lambda ~
1900 $\AA$ for T$\sim$6 K. Experiments in higher fields (e.g. 200
Oe) clearly show a triangular vortex lattice in both real space
(13 $\mu$m by 13 $\mu$m Bitter decoration image of over 1000
vortices) and reciprocal space.
\end{abstract}

\pacs{74.60.Ge, 74.70.Ad}

\maketitle

The recent discovery of superconductivity with $T_c \approx$ 39 K
in the simple intermetallic compound, magnesium diboride,
\cite{Jap} has caused an explosion of experimental and theoretical
works with a major part of the measurements being performed on
polycrystalline samples. \cite{P1,P2,P3,P4,P5}  Until now only a
few groups have been able to grow (small, sub-mm size) single
crystals of MgB$_2$. \cite{X1,X2,X3,X4} From the very earliest
data \cite{P2} it became clear that MgB$_2$ is a type-II
superconductor with its electromagnetic properties described
within the framework of the vortex state. \cite{AAA} So far, for
the most part, bulk techniques (magnetization, magneto-transport,
etc.) that evaluate the "average" properties of the sample were
used for studies of the superconducting state of MgB$_2$. On the
other hand, direct imaging of the vortices is more of a local
probe that can evaluate the homogeneity and strength of pinning
for different parts of the sample. In addition this technique
allows for the determination of basic superconducting properties
such as anisotropy and London penetration depth $\lambda$.

In this work we use one of the direct techniques for the
imaging of vortices:  high resolution Bitter decoration, a technique that
allows for the observation of the individual vortices (in small applied
magnetic field) as well as for the imaging of vortex structures in a wide range of magnetic fields (up to 2kOe \cite{PRBLu}). It should be emphasized, though,
that this technique requires that the surface of the sample be very
clean and optically smooth.  To achieve this degree of surface perfection, single crystals have to be used.

The single crystals used for our decoration experiments were grown
using a high pressure cubic anvil technique from a mixture of Mg
and B in a BN container (see \cite{X4,X5} for details). The
samples used for decoration were plates with the approximate
dimensions $0.4 \times 0.6 \times 0.05 mm^3$. In addition to pure
MgB$_2$ single crystals ($T_c = 38.4$ K, $\Delta T_c = 0.9$ K),
crystals of nominal composition Mg$_{0.99}$Al$_{0.01}$B$_2$ and
Mg$_{0.98}$Al$_{0.02}$B$_2$ (both with $T_c = 35.6$ K, $\Delta T_c
= 0.6$ K) were studied. The decoration was performed in the
field-cooled (frozen flux) regime in applied magnetic fields ($H
\parallel c$) from several Oersted to 200 Oe. The temperature of
the sample before the decoration was either 1.4 K or 4.2 K, during
the decoration process the temperature increased several degrees;
up to 4-5 K or 6-8 K respectively by the end of the decoration
process. The vortex structures were observed on the as grown
surfaces of the crystals using field emission scanning electron
microscope in the secondary electron emission regime to locate the
small islands of iron.
\begin{figure}[b]
 \includegraphics[width=0.8\linewidth]{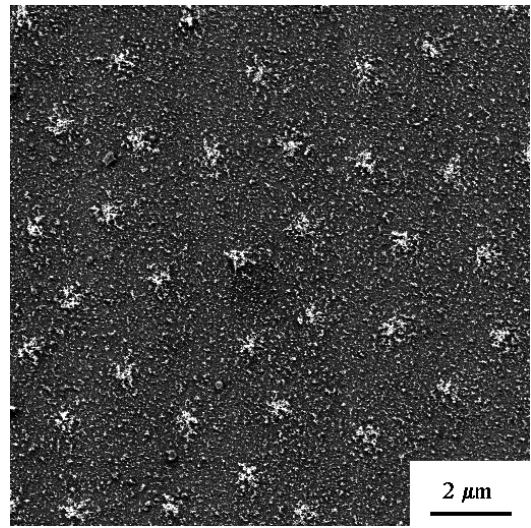}
 \caption{SEM imaging of the vortices at small magnetic field in MgB$_2$ single crystal.\label{1}}
 \end{figure}

Figure 1 shows the structure of vortices in MgB$_2$ for a small
applied magnetic field ($B \approx 4.4$ G). The observed structure
is a collection of weakly interacting individual vortices that do
not form a regular triangular lattice (and do not have long range
order). The diameter, d, of the image of a single vortex ("vortex
diameter") is, on average, 0.77$\pm$0.2 $\mu$m, a value that is
much less than the distance between vortices. Taking into account
the vortex expansion \cite{Pearl} near the surface of a
superconductor, the London penetration depth can be estimated as
$\lambda = kd$ with $k \approx 1/4$ \cite{lambda} that results in
$\lambda \approx 1900 \AA$ for the temperature of the decoration
experiment $T \approx $ 6K. Common techniques for the measurements
of London penetration depth in superconductors usually give very
precise \textit{relative changes} of $\lambda$ as a function of
temperature and/or applied magnetic field. At the same time the
accuracy in the \textit{absolute value} of $\lambda$ is usually
around several tens-of-percents. Since the range of the
experimental values of the penetration depth for MgB$_2$ obtained
using different techniques is rather wide ($\approx 600-3000 \AA$
\cite{blah5,5a,5b,5c}) our estimate of $\lambda$ from the size of
the vortex image is useful by virtue of giving a reliable absolute
value to the upper limit of the penetration depth. Several issues
should be remembered in the course of such estimate. It is
important to have the density of magnetic particles high enough to
fill the region of the magnetic flux penetration in the vicinity
of the vortex. From Fig. 1 it is clear that this condition is
satisfied since the magnetic particles are observed in the area
between the vortices. The correct estimate of the value of the
coefficient $k$ that accounts for the vortex expansion near the
surface of a superconductor is apparently the main source of
uncertainty in our estimate of $\lambda$. Empirically, a
comparison between the values for the penetration depth from
decoration experiments and those obtained by other techniques in
different materials where the values of $\lambda$ are considered
to be reliable (Nb, $\rm YBa_{2}Cu_{3}O_{7-\delta}$, NbSe$_2$)
\cite{vin1,vin2} give the upper limit of errors in $\lambda$ from
Bitter decoration of $\approx 30 \%$ using $k = 1/4$.

\begin{figure}[b]
 \includegraphics[width=0.8\linewidth]{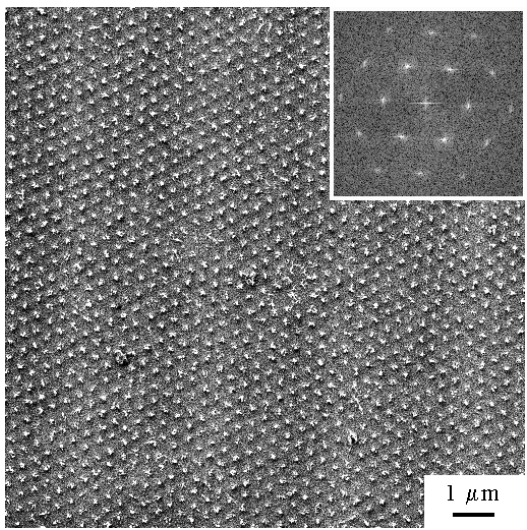}
 \caption{Triangular vortex lattice at magnetic field 200 Oe in MgB$_2$
single crystal, inset: FFT pattern in an arbitrary
scale.\label{2}}
 \end{figure}

In higher applied magnetic fields a regular triangular lattice is
clearly observed (see Fig. 2). No difference in vortex structure
was seen between pure and Al-doped MgB$_2$ crystals. Fig. 2 shows
the vortex lattice for MgB$_2$  single crystal in 200 Oe applied
field. The remarkably high quality of the vortex lattice can be
seen in the real-space image as well as from the fast Fourier
transform (FFT) pattern (inset fig.2). Autocorrelation function
allows to estimate the translational length as 8-9 intervortex
spacings. Our observation of the hexagonal vortex lattice is in a
good agreement with the imaging of vortex unit cell by STM at the
magnetic fields of 2kG \cite{X6} as well as 5 kGs. \cite{X5} It is
worth noting that, given the high degree of order shown by the
flux line lattice in Fig. 2, MgB$_2$ should prove to be an
excellent system for small angle neutron scattering (SANS)
measurements.

\begin{figure}[h]
 \vspace{5mm}
 \includegraphics[width=0.8\linewidth]{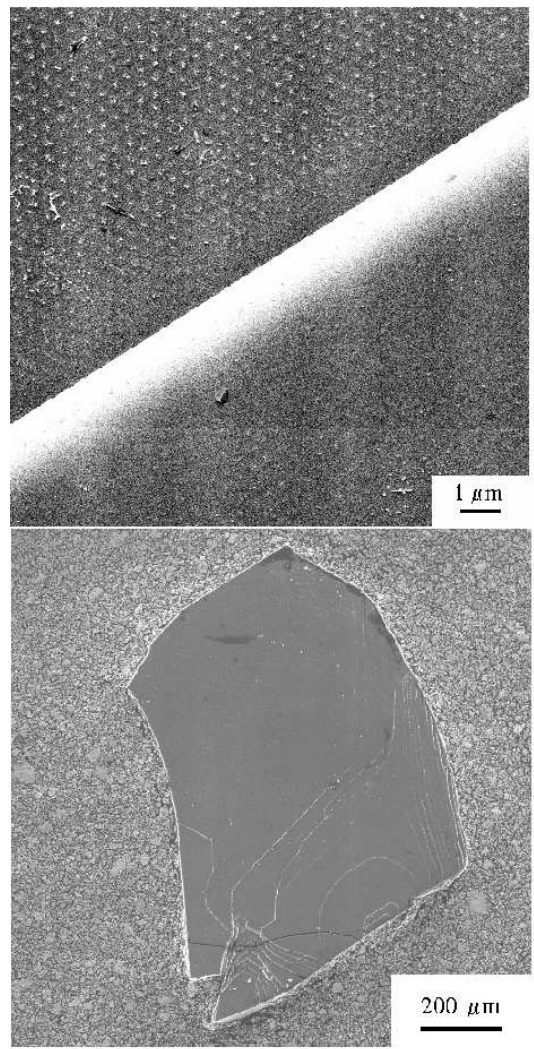}
 \caption{ The vortex lattice in vicinity of the step at magnetic field 200 Oe (top panel),
  full image of the surface of the MgB$_2$ single crystal (bottom panel ).
   Note that the step shown in (top panel) can bee seen in the lower half of (top panel) making an ~ 45 degree angle to the horizontal.
\label{3}}
 \end{figure}

In pure and doped magnesium diboride single crystals Meissner rims
(narrow regions free from vortices) were often observed (see Fig.
3) near the top of growth steps and crystal edges. In the latter
case the width of the stripe was several tens of microns that is
comparable with the thickness of the crystals. Meissner rims were
observed in several other superconductors \cite{blah,blah1,blah2}
with weak volume pinning in which case interaction of vortices
with the lateral surface of the crystal becomes significant. The
observation of a regular triangular lattice and Meissner rims in
pure and Al-doped magnesium diboride single crystals point to weak
volume pinning and therefore to high quality, with regards to
pinning, of the crystals used in the decoration experiments.

As part of our search for effects caused by the expected anisotropy of the London
penetration depth \cite{blah3} we made several attempts to perform
Bitter decoration experiments in a tilted magnetic field using the
same single crystals that were utilized for $H \parallel c$
measurements shown above (after cleaning the samples from magnetic particles).
However imperfections of the observed vortex structure (smeared
maxima in the FFT pattern) did not allow us to reach any unambiguous
conclusions about the anisotropy of $\lambda$. These imperfections
could be caused, at least in part, by possible damage of the
surface of the crystals during the cleaning process.
Decoration experiments in tilted and perpendicular to $c$ magnetic
field will be part of the future studies and still do have the potential of
addressing important issues of superconducting anisotropies in
MgB$_2$.

In summary, a clear triangular vortex lattice was observed in
an applied field of $\simeq 200$ Oe ($H \parallel c$) for
Mg$_{1-x}$Al$_x$B$_2$ single crystals ($x$ = 0, 0.01, 0.02) by
Bitter decoration technique. The Meissner rims seen near the growth
steps and crystal edges suggest very small volume pinning in these
crystals. And an upper limit of the London penetration depth $\lambda \approx
1900 \AA$ at T $\approx$ 6 K was estimated from decoration
experiments in very low fields.

%

\begin{acknowledgments}
We wish to thank V.G. Kogan for useful discussions. Ames
Laboratory is operated for the U. S. Department of Energy by Iowa
State University under Contract No. W-7405-Eng.-82. This work was
supported by the director for Energy Research, Office of Basic
Energy Sciences. Partial support by NATO Collaborative Linkage
Grant \#PST.CLG.978513 is acknowledged. L.Ya.V. was partially
supported by RFBR 00-02-04019a grant and Minestry of
Industry,Science and Technologies grant 40.012.1.1.11.46 (Russia).
\end{acknowledgments}


\end{document}